\documentclass{amsart} 
\usepackage{graphicx}
\usepackage{amssymb, amsmath, hyperref, color}
\usepackage{amsfonts}
\usepackage{amssymb}
\usepackage{float}
\newtheorem{theorem}{Theorem}

\newtheorem{definition}[theorem]{Definition}

\newcommand{\divv}{\text{\rm div}}

\def\f{\frac}
\def\p{\partial}
\def\be{\begin{equation}}
\def\ee{\end{equation}}

\newcommand{\C}{\mathbb C}
\newcommand{\N}{\mathbb N}

\newcommand{\R}{\mathbb R}

\numberwithin{equation}{section}
\numberwithin{theorem}{section}
\numberwithin{example}{section}
\numberwithin{figure}{section}

\def\p{\partial}

\begin{document}
\bibliographystyle{siam}

\title[Geophysical Fluid Dynamics and Climate Dynamics]{Some Mathematical and Numerical Issues in Geophysical Fluid Dynamics  and Climate Dynamics}

\author[Li]{Jianping Li}
\address[JL]{State Key Laboratory of Numerical Modeling 
for Atmospheric Sciences and Geophysical Fluid Dynamics (LASG),
Institute of Atmospheric Physics (IAP), Chinese Academy of Sciences (CAS), 
P.O. Box 9804, Beijing 100029, China}

\author[Wang]{Shouhong Wang}
\address[SW]{Department of Mathematics,
Indiana University, Bloomington, IN 47405\newline
\href{http://www.indiana.edu/~fluid}{\textcolor{blue}{http://www.indiana.edu/\~{ }fluid}}}
\email{showang@indiana.edu}

\thanks{The authors would like to thank two anonymous referees for their insightful comments.
The work was supported in part by the
grants from the Office of Naval Research, from 
the National Science Foundation, and from the National Nature Science Foundation of China (40325015).}


\begin{abstract}
In this article, we address both recent advances and open questions  in  some mathematical and computational issues in geophysical fluid dynamics (GFD) and climate dynamics.  The main focus is on  1) the primitive equations (PEs) models and their related mathematical and computational issues, 2)  climate variability, predictability  and successive bifurcation, and 3)   a new dynamical systems theory and its applications to GFD and climate dynamics.
\end{abstract}
\maketitle
\tableofcontents

\section{Introduction}
\label{sc1}
The atmosphere and ocean around the earth are rotating geophysical fluids, which are also two important components of the climate system.  The phenomena of the atmosphere and ocean are extremely rich in their organization and complexity, and a lot of them cannot be produced by laboratory experiments.   The atmosphere or the ocean or the couple atmosphere and ocean can be viewed as an initial and boundary value problem (Bjerknes \cite{bjerknes}, Rossby \cite{rossby26}, Phillips \cite{phillips}), or an infinite dimensional dynamical system. These phenomena involve a broad range of  temporal and spatial scales (Charney \cite{charney48}). For example, according to J. von Neumann \cite{neumann}, the motion of the atmosphere can be divided  into three categories depending on the time scale of the prediction. They are motions corresponding respectively to the short time, medium range and long term  behavior of the atmosphere.  The understanding of these complicated and scientific issues necessitate a joint effort of scientists in many fields.  
Also, as John von Neumann \cite{neumann} pointed out, 
this difficult problem  involves a combination of modeling, mathematical theory and  scientific computing.

%Atmosphere sciences have been, for a long time, the source of new developments 
%in the mathematical and other sciences. The most recent and notable one may 
%be the introduction of the Lorentz attractor, but there are also numerous 
%important developments stemming from meteorology: for example the interest 
%and the impact of Richardson on meteorology and computing in the early 1920's,
%or the work of John von Neumann, J. Charney in this area, 
%or the introduction by 
%Smagorinsky of his eddy-viscosity model which has been the basis for 
%many developments in computational fluid mechanics (large eddy simulations). 
%At this time the study of global climate problems is the source of some 
%of the most extensive calculations and it produces one of the most 
%significant fields for testing the largest and most powerful computers. 

In this article, we shall address mathematical and numerical issues in geophysical fluid dynamics and climate dynamics. The main topics include
\begin{enumerate}

\item issues on the modeling, mathematical analysis and numerical analysis of the primitive equation (PEs), 

\item climate variability, predictability  and successive bifurcation,

\item a new dynamical systems theory  and  its applications to geophysical fluid dynamics.

\end{enumerate}

%The first is on the general  models for large scale atmosphere and ocean,
% and the  second is on
%One objective  is to study some theoretical  issues related to the primitive equations (PEs).  
As we know, the atmosphere is a compressible fluid and the seawater is a slightly compressible fluid. The governing equations for either the atmosphere, or the ocean, or the coupled atmosphere-ocean models  are the general equations of hydrodynamic equations together with other conservation laws for such quantities as the energy, humidity and salinity, and with proper boundary and interface conditions. Most general circulation models (GCMs) are based on the PEs,  which are derived using the hydrostatic assumption in the vertical direction. This assumption is due to the smallness of the aspect ratio (between the vertical and horizontal length scales). 
We shall present a brief survey on recent theoretical and computational developments and future studies of  the PEs.

One of the primary goals in climate dynamics is to document, through careful 
theoretical and numerical  studies, 
the presence of climate low frequency variability, 
to verify the robustness of this variability's characteristics to 
changes in model parameters, and to help explain its physical mechanisms. 
The thorough understanding of this variability is a 
challenging problem with important practical implications 
for geophysical efforts to quantify predictability, analyze 
error growth in dynamical models,
and develop efficient forecast methods.
As examples, we discuss a few sources of variability, including
wind-driven (horizontal) and thermohaline (vertical) circulations, 
El Ni\~no-Southern Oscillation (ENSO),  and Intraseasonal oscillations (ISO). 

The study of the above geophysical problems  involves on the one hand applications of the existing  mathematical and computational  theories to the understanding of the underlying physical problems, and on the other hand the development of new 
mathematical theories. 

We shall present briefly a dynamic bifurcation and stability theory  and its  applications to GFD. This theory, developed recently by  Ma and Wang \cite{b-book},  is for both finite and infinite dimensional
dynamical systems, and  is centered at a new notion of bifurcation, called attractor bifurcation. The theory is briefly described by a simple system of two ordinary differential equations, and by the classical Rayleigh-B\'enard convection. Applications to 
the stratified Boussinesq equations model and the doubly-diffusive models are also addressed.  

We would like to mention that there are many important issues not covered in this article, including, for example, the  ocean and atmosphere data assimilation and prediction problems, and the  stochastic-dynamics studies; see, among many others, 
\cite{GCTBI, 
GMR, ghil97, GIBCKS, KFIG, KSGT,GR02,GADIKMRS,KKG,KKRG,KKG06,DBDG,KSBDG,
majda-wang} and the references therein.

The article is organized as follows. In Section 2, some basic GFD models are introduced, with some mathematical and computational issues given in Section 3. Section 4 is on predictability, and Section 5 deals with issues on climate variability. Section 6
presents the new  dynamical systems theory based on attractor bifurcation and its application to Rayleigh-B\'enard convection and to GFD models. 

 \section{Modeling}
\subsection{The primitive equations (PEs) of the atmosphere}
%The motion of the atmosphere has been of interest to scientists for a 
%long time, with early contributions from L. da Vinci, E. Halley, 
%G. Hadley, P. Laplace and 
%then the english scientist L. F. Richardson, the norwegian scientist 
%, swedish meteorologist C. G. Rossby.
%Finally in the late 1940's and the early 1950's, with the appearance of 
%computers, J. von Neumann, J. Charney and their collaborators 
%have systematically  introduced the scientific methods in meteorology 
%and based the prediction of the weather and climate  on the 
%numerical simulation of the partial differential equations which 
%describe the motion of the atmosphere (supplementing on the other hand 
%the accumulation of data).
%Nowadays the prediction of weather and the study  of the atmosphere is 
%the every day preoccupation of many laboratories including NCAR  and those 
%involved with aviation, space or the air pollution in the large cities.

Physical laws governing the motion and states of the 
atmosphere and ocean can be described by  the general equations of  
hydrodynamics and thermodynamics. 
Using a non-inertial coordinate system rotating with the earth, 
these equations can be written as follows:
\begin{equation}
\left.
\begin{aligned}
& \frac{\p \vec V}{\p t} + \vec V \cdot \nabla_3 \vec V 
+ 2 \vec \Omega \times \vec V -\vec g + \frac1{\rho}\text{grad}_{3} p 
=\vec D_M,\\
& \f{\p  \rho}{\p t} +  \text{div}_3  (\rho \vec V)=0,\\
& c_v \f{\p  T}{\p t} + c_v \vec V \cdot \nabla_3 T 
+ \frac{p}{\rho} \nabla_3 \vec V=Q + D_H,\\
& \f{\p  q}{\p t} +\vec V \cdot \nabla_3 q=\f{S}{\rho}+ D_q, \\
& p= R\rho T.
\end{aligned}\right. \label{original}
\end{equation}
Here the first equation is the momentum equation, the second is the continuity equation, the third is first law of thermodynamics, the fourth is diffusion equation for the humidity, and the last is the equation of state for ideal gas. The unknown functions are  the three dimensional velocity field $\vec V$, the density function $\rho$, 
the pressure function $p$, the  temperature function $T$, and the specific humidity function 
$q$. Moreover, in the above equations,  $\vec \Omega$ stands for the angular velocity of the earth,  $\vec g$  the gravity, $R$  the gas constant,  $c_v$  the specific heat at constant volume,
$\vec D_M$   the viscosity terms, $D_H$   the temperature diffusion, 
$Q$   the heat flux per unit 
density at the unit time interval, which includes molecule or turbulent, 
radiative  and evaporative heating,  and $S$ the differences of the rates of the evaporation and condensation.

These equations are normally  far too complicated; 
simplifications from both the physical and mathematical 
points of view are necessary. There are essentially two 
characteristics of both the atmosphere and ocean, which are  used in 
simplifying the equations. The first one is that for large scale 
geophysical flows, the ratio between the vertical and horizontal scales 
is very small; this leads to the primitive equations (PEs) of
 both the atmosphere and the ocean, which are the basic equations for  
these two fluids. More precisely, the PEs are obtained from 
the general equations of  hydrodynamics and thermodynamics 
of the compressible atmosphere, by approximating the momentum 
equation in the vertical direction with the hydrostatic equation:
\begin{equation}\label{2.2}
\frac{\p p}{\p z}= -\rho g. 
\end{equation}
This hydrostatic equation is 
based on the ratio between the vertical and horizontal scale 
being small.
Here $\rho$  is the density, $g$ the gravitational constant, 
and $z=r-a$  height above the sea level, $r$ the radial 
distance, and $a$ the mean radius of the earth. 
Equation (\ref{2.2}) expresses the fact that $p$ is a decreasing
function along the vertical so that one can use $p$ instead of $z$ as
the vertical variable.  
Motivated by this hydrostatic approximation, we can introduce a 
generalized vertical coordinate system $s$-system given by 
\begin{equation}
s=s(\theta, \varphi, z, t),
\end{equation}
where  $s$ is a strict monotonic function of $z$. 
Then the basic equations of the large-scale atmospheric motion 
in the s-system are
\begin{equation}
\left.
\begin{aligned}
& \frac{\partial v}{\p t} + v \cdot \nabla_s v + {\dot s} \f{\p v}{\p s} 
+ f k\times v + \frac{1}{\rho} \nabla_s z =D_{M}, \\
& \f{\p p}{\p s}\f{\p s}{\p z} + \rho g=0, \\
& \f{\p}{\p s}\left(\f{\p p}{\p t}\right)_s + 
\nabla_s \cdot \left(v \f{\p p}{\p s}\right) + \f{\p}{\p s}\left(\dot s 
\f{\p p}{\p s}\right)=0, \\
& c_v \frac{\p T}{\p t} + c_v v \cdot \nabla_s T 
+ c_v \dot s \f{\p T}{\p s} 
+ \frac{1}{\rho} 
\left(\frac{\p p}{\p t} + v \cdot \nabla_s p 
+ \dot s \f{\p p}{\p s} \right)
= Q + D_H, \\
& \frac{\p q}{\p t} + v \cdot \nabla_s q 
+ \dot s \f{\p q}{\p s} = \frac{S}{\rho} + D_q.
\end{aligned}
\right.
\end{equation}
Some common $s$-systems in meteorology are respectively the $p$-system 
(the pressure coordinate), the $\sigma$-system 
(the transformed pressure coordinate), 
the $\theta$-system (the isentropic coordinate), and 
the $\zeta$-system 
(the topographic coordinate or transformed height coordinate).
The above PEs appear in the
literature in e.g. the books of A.~E. Gill \cite{gill}, G. Haltiner and R.
Williams \cite{HW80}, J. R. Holten \cite{holden}, J. Pedlosky \cite{Pb}, 
J.~P. Peixoto and A.~H. Oort \cite{Peixo},
W.~M. Washington and C.~L. Parkinson \cite{WP86}, Q. C. Zeng \cite{zeng}.  
We remark here that sometimes the pressure coordinate is  denoted by $\eta$, and  terrain-following by  $\sigma$.

For simplicity, here we discuss only the case with 
the coordinate transformation
from $(\theta, \varphi, z)$ to $(\theta, \varphi, p)$.
The basic equations of the atmosphere are  
then the {\it Primitive Equations  (PEs)} of the atmosphere in the 
$p$-coordinate system. 
As they appear in classical meteorology books (see 
e.g. \cite{zeng}  and   Salby \cite{Salb96}),  the PEs are given by 
\begin{equation} \label{2.1}
\left.
 \begin{aligned}
&{\partial v\over\partial t}+   v\cdot \nabla v+\omega
        {\partial v\over\partial p}+2\Omega\cos \theta 
            k\times v+\nabla \Phi  =D_M ,\\
&{\partial\Phi\over\partial p}+{{RT}\over p}=0,\\
&\hbox{div }v +{\partial\omega\over\partial p}=0,\\
& {\partial T\over\partial t}
   +v \cdot \nabla T +\omega\left({\kappa T \over p}
  -   {\partial T\over\partial p}\right)  =
        \frac{\tilde Q_{rad}}{c_p}+  
      \frac{\tilde Q_{con}}{c_p} + D_H ,\\ 
&{\partial q\over\partial t}+v \cdot \nabla q
   +\omega {\partial q\over\partial p}
          =E-C + D_q,
\end{aligned} \right.
\end{equation}
where $D_M$ is the dissipation terms for momentum and $D_H$
and $D_q$ are diffusion terms for heat and moisture, respectively, 
$E$  and $C$ are the rates of evaporation and condensation due 
to cloud processes, $c_p$  the heat capacity, and $\tilde Q_{rad}$
and $\tilde Q_{con}$ the net radiative heating and the heating due 
to condensation processes, respectively. 
We use  the pressure coordinate system 
$(\theta, \varphi, p)$, where $\theta$ ($0 < \theta < \pi$) 
and  $\varphi$ ($0 < \varphi < 2\pi$)  are 
the  colatitude and longitude  variables, and  $p$ the pressure of the air. 
The nondynamical processes 
$\tilde Q_{rad}$, $\tilde Q_{con}$, $E$  and $C$  are called model physics.
Furthermore, the unknown functions are the horizontal velocity $v$, 
the vertical velocity $\omega=dp/dt$, the geopotential $\Phi$, 
 the temperature $T$, and 
specific humidity $q$. The operators $\hbox{div}$ and $\nabla$ 
are the two dimensional operators on the sphere.

Of course, this set of equations is supplemented with  a set of 
physically sound boundary conditions  such as  (\ref{3bc}), depending 
on the specific form of the forcing and dissipation.

\subsection{Ocean  models}
The sea water is almost an incompressible fluid, leading to the Boussinesq approximation, i.e., a variable density is only recognized  in the buoyancy term and the equation of state. The resulting equations are called the Boussinesq equations given as follows:
\begin{equation}\label{oceanpe}
\left.
\begin{aligned}
& {\partial v\over \partial t}+\nabla_v v +w{\partial v\over \partial
z}+ {1\over\rho_0}\hbox{grad}\rho+fk\times v -\mu\triangle v-\nu{\partial ^2v \over\partial
z^2}=0,\\
&  {\partial w\over\partial t}+\nabla_v w +w{\partial
w\over\partial z}+{1\over\rho_0}{\partial\rho \over \partial
z}+{\rho\over \rho_0}g -\mu\triangle w-\nu{\partial^2w\over\partial z^2}=0,\\
& \text{div}v+{\partial w\over\partial z}=0,\\
& {\partial T\over\partial t}+\nabla_vT+w{\partial T\over\partial
z}-\mu_T\triangle T-\nu_T{\partial^2T\over\partial
z^2}=0,
\\
&  {\partial S\over \partial t}+\nabla_vS+w{\partial S\over \partial
z}-\mu_S \triangle S-\nu_S{\partial ^2S\over \partial
z^2}=0,\\
& \rho=\rho_0(1-\beta_T(T-\bar T_0)+\beta_S(S-\bar S_0)),
\end{aligned}\right.
\end{equation}
where $v$ is the horizontal velocity field,  $w$ the vertical
velocity, and $S$ the salinity.  The sixth equation in (\ref{oceanpe})  is an empirical equation for the density function based on the linear approximation. In general, density $\rho$ is a nonlinear function of $T$, $S$, and $p$. With higher approximations, one will encounter additional mathematical difficulties although the nonlinear equation of state is essential for some elements of ocean circulation (e.g., cabbeling). 
%From the physical point of view, however, the need of higher order 
%approximations is not clearly justified.

As in the atmospheric case, the hydrostatic assumption is usually used, leading to the PEs for the large-scale ocean:
\begin{equation}
\left.
\begin{aligned}
&{\partial v\over\partial t} +\nabla_{v}v+w
{\partial v\over\partial z}+
{1\over\rho_0}\nabla p
+f  k\times v-\mu_v\Delta_a v-
\nu_v{\partial^2v\over\partial z^2}=0,\\
&{\partial p\over\partial z}=-\rho g,\\
&\hbox{ div }v+{\partial w\over\partial z}=0,\\
&{\partial T\over\partial t}+\nabla_{v}T+w
{\partial T\over\partial z}
-\mu_T\Delta_a T-\nu_T
{\partial^2T\over\partial z^2}=0,\\
&{\partial S\over\partial t}+\nabla_{v}S+
w{\partial S\over\partial z}-\mu_s\Delta_a S-
\nu{\partial^2S\over\partial z^{2}}=0,\\
&\rho=\rho_0
(1-\beta_T(T-\bar T_0)+\beta_S(S-\bar S_0)).
\end{aligned}\right.
\end{equation}
Also, we note that if the hydrostatic assumption is made first, the  
Boussinesq approximation is not really necessary (see eq. (2.5) -  
divergence-free! - or, e.g., de Szoeke and Samelson \cite{SSamelson}.

\subsection{Coupled atmosphere-ocean models}
Coupled atmosphere and ocean models  consist of 1) models for the atmosphere component, 2)  models for the ocean component, and 3) interface conditions. 
The interface conditions are used to couple the atmosphere and ocean systems, and are usually derived based on on first principles; see Gill \cite{gill}, Washington and Parkinson \cite{WP86}. A mathematically well-posed coupled model with physically sound interface conditions and the PEs of the atmosphere and the ocean are given in Lions, Temam and Wang \cite{LTW93}.  We refer interested readers to these references to further studies. 

As we know, the atmosphere and ocean components have quite different time scales, leading to complicated dynamics. For example, from the computational point of view, one needs to incorporate the two time scales; see e.g. \cite{LTW96}. 

\section{Some Theoretical and Computational Issues for the PEs}

\subsection{Dynamical systems perspective  of the models}
From the mathematical point of view, 
we can put the models addressed in the previous section in the 
perspective of infinite dimensional dynamical systems as follows:
\begin{equation}
\label{3.1}
\left. \aligned
& \varphi_t + A\varphi + R(\varphi) = F, \\ 
& \varphi|_{t=0} = \varphi_0,
\endaligned \right. 
\end{equation}
defined on an infinite dimensional phase space $H$.
Here $A: H \to H$ is an unbounded linear operator, $R:H \to H$ is a 
nonlinear operator, $F$ is the forcing term, 
and $\varphi_0$ is the initial data.

We remark here that the linear operator can usually be written as 
$A=A_1+A_2$, where $A_1$ stands for 
the irreversible diabatic linear processes of energy dissipation, and 
$A_2$ for the reversible adiabatic linear processes of energy conversation. 
The nonlinear term 
$R(\varphi)$ represents the 
reversible adiabatic nonlinear processes of energy conversation.
The properties of these operators
reflect directly  
the essential characteristics of two kinds of basic 
processes with entirely different physical meanings.

The above formulation 
is often achieved by (a) establishing a proper functional setting of 
the model, and (b) proving the existence and uniqueness of the solutions.

%\subsection{PEs as a dynamical system}
Hereafter we demonstrate the procedure with the PEs. 
Due to some  technical reasons, 
some minor and physically reasonable modifications of the PEs are made. 
In particular, we assume that the model physics 
$\tilde Q_{rad}$, $\tilde Q_{con}$, $E$  and $C$
are given functions of location and time. 
We specify also the viscosity, diffusion terms as 
(see among others Lions {\it et. al.} \cite{LTW92a} and Chou \cite{choujf}):
\begin{equation}\label{3.2}
\left.
 \begin{aligned}
& D_M=- L_1 v, \\ 
& \frac{\tilde Q_{rad}}{c_p}+  
      \frac{\tilde Q_{con}}{c_p} + D_H =- L_2 T + Q_T, \\
&  E-C + D_q =- L_3 q + Q_q, \\
& L_i=-\mu_i \Delta 
  - \nu_i \frac{\partial}{\partial p}
         \left(\left(\frac{gp}{R\bar T(p)}\right)^2 
      \frac{\partial }{\partial p}\right), 
\end{aligned} \right.\end{equation}
 where $\mu_{i},\nu_{i}$ are horizontal and vertical
viscosity and diffusion coefficients,  $\Delta$ is 
the Laplace operator
 on the sphere, $Q_T$ and $Q_q$ are treated as given functions, 
and $\bar T(p)$  a 
given temperature profile, which can be considered as the climate average 
of $T$.
The boundary conditions for the PEs are given by 
\begin{equation}
\left.
\begin{aligned}
& \f{\partial }{\partial p}(v, T, q)
=(\tilde \gamma_s(\tilde v_s-v), \tilde \alpha_s(\tilde T_s-T), 
\tilde \alpha_q(\tilde q_s-q) ), \ 
     \omega=0 &&  \text{at }  p=P,\\
&\f{\partial }{\partial p}(v, T, q)=0, \ \omega=0
      &&  \text{at } p=p_0.
\end{aligned}\right. \label{3bc}
\end{equation}

The second and third equations (\ref{2.1}) are diagnostic 
ones; integrating  them in $p$-direction, we obtain 
\begin{equation}\label{3.3}
\left.
\begin{aligned}
&\int^P_{p_{0}}\hbox{ div }v(p^\prime)dp^\prime =0, \\
&\omega =W(v)=-\int^p_{p_{0}}\hbox{ div }v(p^\prime)dp^\prime, \\
&\Phi =\Phi_s+ \int^P_p
\displaystyle{{RT(p^\prime)\over p^\prime}}dp^\prime. \\
\end{aligned} \right. \end{equation}
Then the PEs are equivalent to the following functional formulation:
\begin{equation}\label{3.4}
\left.
\begin{aligned}
&\displaystyle{{\partial u\over\partial t}}+\Lambda(v)u+P(u)+Lu + 
( \nabla \Phi_s, 0, 0) =(0, Q_T,  Q_q), \\
& \text{div }\int^P_{p_0} v dp =0,  \\
\end{aligned} \right. \end{equation}
where $ u=(v,T, q) $,  $\Lambda(v)u=\nabla_v u 
+ W(v)\displaystyle{{\partial u \over\partial p}}$,  
$Lu$  corresponds to the viscosity and diffusion terms,
and $Pu$ the lower order terms. 
The above new formulation was first introduced by Lions, Temam and Wang 
\cite{LTW92a}. 

We solve then the PEs in some 
infinite dimensional phase spaces $H$ and $V$. In 
particular, we use 
$$H_1 =\left\{v  \in L^2 \;\; | \;\; 
   \text{div } \int^P_{p_0} v dz =0 \;\; \right\}$$
as the phase space for the horizontal velocity $v$. 
Then we project in the phase space. Using this projection, the unknown 
function $\Phi_s$ plays a role as a Lagrangian multiplier, 
which can be recovered by the following decomposition:
\begin{align*}
& L^2= H_1 \oplus H^{\perp}_1, \\
& H^{\perp}_1 =\left\{ v \in L^2 \;\; | \;\; v=\nabla \Phi_s, 
   \Phi_s \in H^1(S^2_a)\right\}.
\end{align*}
Then the PEs are equivalent to an infinite dimensional dynamical system 
as (\ref{3.1}). 

With the above formulation, for example, 
we encounter the following new nonlocal Stokes problem:
\begin{equation} \label{3.5}
\left.
\begin{aligned}
& - \triangle v + \nabla \Phi_s = f, \\
& \text{div }\int^P_{p_0} v dp =0, \\
\end{aligned}\right. 
\end{equation}
supplemented with suitable boundary conditions. 
From the mathematical point of view, all techniques 
for the regularity of solutions are local. But our problem here is 
nonlocal;  the regularity of the solutions for this problem  
can be obtained using Nirenberg's finite difference quotient method; see 
\cite{LTW92a}.

%The above new formulation was first introduced by Lions, Temam and Wang 
%\cite{LTW92a}. Many theoretical work has been done, notably by several groups 
%including a) Lions, Temam and Wang and their collaborators, 
%%b) J. Chou, Jianping Li and Q. C. Zeng and their collaborators, 
%c) french schools, ...

%\subsection{Other models}
Other models such as the PEs of the ocean and the coupled atmosphere-ocean 
models can be viewed as infinite dimensional dynamical systems 
in the form of  (\ref{3.1}). We refer the interested readers to 
Lions, Temam and Wang \cite{LTW92b} for the PEs of the ocean, and 
 \cite{LTW93,LTW95} for the coupled atmosphere-ocean models.

\subsection{Well-posedness}
One of the first mathematical questions is the existence, 
uniqueness and regularity  of solutions of the 
models. The main results in this direction 
can be briefly summarized as follows, and we refer the interested readers to 
the related references given below for more details: 

For the PEs of the large-scale atmosphere, 
the existence of global (in time) 
weak solutions of the primitive equations for 
the atmosphere is obtained by Lions, Temam and Wang \cite{LTW92a}, 
where the new formulation described above is introduced.

In fact, the key ingredient for most, if not all,  existence results for the PEs of the atmosphere-only, the ocean-only, or the coupled atmosphere-ocean  depend heavily on the formulation (\ref{3.3}) and (\ref{3.4}), first introduced by  Lions, Temam and Wang 
\cite{LTW92a}. 
We note that without using this new formulation, one can also 
obtain the existence of global weak solutions by introducing 
proper function spaces with more regularity in the $p$-direction; see 
Wang \cite{Wang88}. However, the new formulation is important 
for viewing the PEs as an infinite dimensional system.

The existence of global strong solutions is first obtained for small data  and large time or for short time by  \cite{Wang88, TZ04}. Other studies include the case for thin domain case \cite{HTZ} (see also Ifitimie and Raugel \cite{raugel} for discussions on the Navier-Stokes equations on thin domains),  and  for fast rotation \cite{BMN}. 

Recently,  for the primitive equations of the ocean with free top and bottom boundary conditions, an existence of  long time strong solutions with  general data  is obtained by Cao and Titi \cite{CT}, using   the new formulation and the fact that the surface pressure depends on the two spatial directions, and then similar result is obtained for the Dirichlet boundary conditions by Kukavica and Ziane \cite{kukavica}. Furthermore, Ju \cite{juning} studied the global attractor for the primitive equations. 

The corresponding results can also be 
obtained for the PEs of the ocean; see \cite{LTW92b,TZ04, CT} and the 
references therein for more details. 
Existence of global weak solutions were obtained for 
the coupled atmosphere-ocean model introduced in \cite{LTW95}. 
%Other studies for the coupled atmosphere-ocean models include 
%regularity of solutions of the corresponding nonlocal coupled atmosphere 
%ocean  Stokes problems include \cite{Z97}.

As mentioned earlier, the hydrostatic assumption and its related formulation given by 
(\ref{3.4})   and (\ref{3.5}) are crucial in many of the existence  results for both strong and weak solutions. We would like to point out that Laprise  \cite{laprise} suggests  that hydrostatic-pressure coordinates could be used advantageously in nonhydrostatic atmospheric models based on the fully compressible equations. We believe that with the Laprise' formulation, one can extend some of the results discussed here to non-hydrostatic cases. 

Another situation where the new formulation (\ref{3.4})   and (\ref{3.5}) does not appear to be available is  related to more complex low boundary conditions. Some partial results for weak solutions are obtained in  \cite{wmzl}, and apparently many related issues are still open.

\subsection{Long-time dynamics and nonlinear adjustment process} 
Regarding the PEs as an infinite dimensional dynamical system, 
the existence and finite dimensionality of the global attractors 
of the PEs with vertical diffusion were explored in 
\cite{LTW92a, LTW92b, LMTW}. 
The finite dimensionality of the global attractor of 
the PEs provides  a mathematical foundation 
that the infinite dynamical system can be described 
by a finite dimensional dynamical system.

As we know all general circulation  models of either the 
atmosphere-only, or the ocean-only, or the coupled ocean-atmosphere 
systems are based on the PEs with more detailed model physics. 
In both these GCM models  and theoretical climate studies, the PEs  
are   often replaced  by a set of truncated  ordinary
differential equations (ODEs), whose asymptotic solution sets, 
called  attractors, can be investigated in the more tractable 
setting of a finite-dimensional 
phase space, without seriously altering the essential dynamics.
It is, however, not known mathematically whether this truncation is 
really reasonable, and moreover, how we can determine {\it a priori} 
which finite-dimensional truncations are sufficient to 
capture the essential features of the atmosphere or the oceans. 
With this objective, nonlinear adjustment process 
associated with the  long-time dynamics of 
the models were carefully conducted in a series of papers by Li and Chou 
\cite{LC96b,LC96a,LC97a, LC97c,LC97b,LC98b,LC98a, LC99b}; 
see also \cite{whc, LMTW}.

An important consequence of the above mentioned results 
for the long-time dynamics is that 
the nonlinear adjustment process of the climate is a
forced, dissipative, nonlinear system to external forcing. This nonlinear adjustment process  is  different from the  adjustments 
in the traditional dynamical meteorology, including   the geostrophic adjustment, rotational adjustment, 
potential vorticity adjustment, static adjustment, etc. 

These traditional adjustments do not appear to be associated with any attracting properties that attractors usually process.
%and its 
%The existence of a global attractor is simply 
%the basic difference between the adjustment of the climate system 
%and the above adjustments. 
It is indicated from the nonlinear adjustment process 
that as  time increases, the information carried by the initial state will  gradually be lost. In addition,  there are three categories 
of time boundary layers (TBL) and the self-similar structure of 
the TBL in the adjustment and evolution processes of the forced, 
dissipative, nonlinear system.

An important question is how to determine the structure of the 
attractors and the distribution of their attractive domains under 
the given conditions of known external forcing. 
Although the information of initial state will be decayed 
as the time increases, it does not mean that the information of initial 
state is not important to long-time dynamics. The quantity of those 
initial values which locate very close to the boundary between two 
different attractive domains are very important since their local 
asymptotic behavior will be quite different due to slight initial error. 
Another open question is how to construct independent orthogonal basis of 
the attractor in practice based on the finite dimensionality of the 
global attractor. 
Two empirical approaches used at present are the Principal Analysis (PC) or Empirical Orthogonal Function (EOF) and Singular Vector Decomposition (SVD) based on the time series of numerical solutions or observational filed data. They are available although they are empirical. Qiu and Chou \cite{QC}  and Wang et al \cite{wlc}
made a valuable attempt to apply the long-time dynamics of the atmosphere 
mentioned above and this two empirical decomposition methods for orthogonal bases of the attractor  to study 4-dimensional data assimilation.

%Our result suggests clearly
%that the required resolution is quite sensitive to the magnitude
%of the effective (or eddy) viscosity, while it appears to be less
%sensitive to the details of the way that the atmosphere is heated.

\subsection{Multiscale asymptotics and simplified models}
As practiced by the earlier workers 
in this field such as J. Charney  and John von Neumann, and from 
the lessons learned by the failure of Richardson's pioneering work, 
one tries to be satisfied with simplified models approximating 
the actual motions to a greater or lesser degree 
instead of attempting to deal with the atmosphere in all its complexity. 
By starting with models incorporating only 
what are thought to be the most important of atmospheric influences, 
and by gradually bringing in others, one is able to proceed 
inductively and thereby 
to avoid the pitfalls inevitably encountered when a great 
many poorly understood factors are introduced all at once. 

%\subsubsection{Geostrophic asymptotics}
One of the  dominant features of 
both the atmosphere and the ocean is the influence of the 
rotation of the earth.
The emphasis on the importance of the rotation effects of the earth
and their study 
should be traced back to the work of P. Laplace in the
eighteenth century. The question of how a fluid  adjusts
in a uniformly rotating system was not completely discussed until the
time of C. G. Rossby \cite{rossby}, when Rossby considered the process of
adjustment to the geostrophic equilibrium. 
This process is now referred to as the {\it Rossby adjustment}.
Roughly speaking the Rossby adjustment process explains why the
atmosphere and ocean are always close to geostrophic equilibrium, for
if any force tries to upset such an equilibrium, the gravitational
restoring force quickly restores a near geostrophic equilibrium.
Later in the 1940's, under the  
famous $\beta$-plane assumption, Charney \cite{charney} introduced 
the quasi-geostrophic (QG) equations for the {\it large-scale} (with 
horizontal scale comparable to $1000$ $km$) mid-latitude atmosphere.
Since then, there have been many studies  from both the physical and numerical
points of view. This model has been the main 
driving force of the much development of theoretical 
meteorology and oceanography.

Mathematically speaking, the QG theory is based on asymptotics in terms of 
a small parameter, called Rossby number $Ro$, a dimensionless number relating the ratio of inertial force to Coriolis force for a given flow of a rotating fluid.
The key idea in the geostrophic asymptotics, leading to the QG equations, is  to approximate  the spherical midlatitude region by the tangent plane, called 
the $\beta$-plane, at the center of 
the region, and to  express
the Coriolis parameter  in terms of the Rossby number. 

Thanks in particular to the 
vision and effort of Professor J. L. Lions, 
there have been extensive mathematical studies. 
As this topic is very well-received and studied by applied mathematicians, 
we do not go into details, and the interested readers 
are refereed to 
(Lions, Temam \& Wang \cite{LTW94,LTW97}, 
Bourgeois \&  Beale \cite{BB}, 
Babin, Mahalov \& Nicolaenko \cite{BMN}, 
Gallagher \cite{gallagher98},  
Embid \& Majda \cite{EM})
and the references therein. 
In addition, planetary geostrophic  equations (PGEs) of ocean have also received a lot of attention recently following the early work by  Samelson, Temam and Wang \cite{STW1, STW2}.  However, many issues are still open. 

As mentioned earlier, many geophysical processes have multiscale 
characteristics. One aspect of the studies requires a careful examination 
of the interactions of multiple temporal and spatial scales. 
A combination of rigorous mathematics and physical modeling together with 
scientific computing  appear to be crucial for the understanding of these 
multiscale physical processes. ENSO and ISO are two such examples 
(see also Section 5).

\subsection{Some computational issues}
On the one hand, we need to develop more efficient numerical methods for general circulation models (GCMs), including atmospheric general circulation model (AGCM), oceanic general circulation model (OGCM) and coupled general circulation model (CGCM), climate system model and earth system model. On the other hand, numerical simulations are used to test the theoretical results obtained, as well as for preliminary exploration of the phenomena apparent in the governing PDEs,  and to obtain guidance on the most interesting directions for theoretical studies. 

Here we simply address some computational issues without discussing physical processes, which are certainly important in developing GCMs.
Two basic discretization schemes commonly used in GCMs are grid-point and spectral  approaches. There are many crucial issues which are not fully resolved, including  
1) spherical geometry and singularity near the polar regions, 2) irregular domains, 
3) multiscale (spatial and temporal) problems involving both fast and slow processes, 
4)  vertical stratification, 5) sub-grid processes,  6)  model bias, and 7) nonhydrostatic models, etc. We note that although many existing models are  
formulated and solved in spherical coordinates, numerical and mathematical difficulties caused by the dependence of the meshes size on the latitude is still not fully resolved.
Furthermore,  high precision computation and very fine resolutions are two tendencies in developing numerical models.

Recently a fast and  efficient spectral method for the PEs of the atmosphere is introduced by Shen and Wang \cite{sw99}, and further studies for this method applied to more practical GCMs are needed. Another paper uses heavily the surface pressure formulation of the PEs for the ocean is given in Samelson et al. \cite{stww}.  Its incorporation to  OGCMs, and its simulation for studying specific oceanic phenomena 
appear to be necessary.

A remarkable, but neglected, problem  is on influences of round-off error on long time numerical integrations since round-off error can cause numerical uncertainty. Owing to the inherent relationship between the two uncertainties due to numerical method and finite precision of computer respectively, a computational uncertainty principle (CUP) could be definitely existed in numerical nonlinear systems 
(Li et al.  \cite{LZC1, LZC2}), which implies a certain limitation to the computational capacity of numerical methods under the inherent property of finite machine precision.  In practice, how to define the optimal time step size and optimal horizontal and vertical resolutions of a numerical model to obtain the best degree of accuracy of numerical solutions and the best simulation and prediction is an important and urgent computational issue.

\section{Nonlinear Error Growth Dynamics and Predictability}
For a complex nonlinear chaotic system such as the atmosphere or the climate the intrinsic randomness in the system sets a theoretical limit to its predictability; see Lorenz \cite{La, L69}. Beyond the predictability limit, the system becomes unpredictable. In the studies of predictability, the Lyapunov stability theory has been used to determine the predictability limit of a nonlinear dynamical system. The Lyapunov exponents give a basic measure of the mean divergence or convergence rates of nearby trajectories on a strange attractor, and therefore may be used to study the mean predictability of chaotic system; see  Eckmann and Ruelle \cite{ER},  Wolf et al. \cite{WSS}  and Fraedrich 
\cite{Fr}. Recently, local or finite-time Lyapunov exponents have been defined for a prescribed finite-time interval to study the local dynamics on an attractor 
Kazantsev  \cite{Ka},  Ziehmann et al \cite{ZS}  and  Yoden and Nomura \cite{YN}.  However, the existing local or finite-time Lyapunov exponents, which are same as the global Lyapunov exponent, are established on the basis of the fact that the initial perturbations are sufficiently small such that the evolution of them can be governed approximately by the tangent linear model (TLM) of the nonlinear model, which essentially belongs to linear error growth dynamics. Clearly, as long as an uncertainty remains infinitesimal in the framework of linear error growth dynamics it cannot pose a limit to predictability. To determine the limit of predictability, any proposed Ôlocal Lyapunov exponentÕ must be defined with the respect to the nonlinear behaviors of nonlinear dynamical systems Lacarra and Talagrand \cite{LT}   and Mu \cite{Mu00}.

In view of the limitations of linear error growth dynamics, it is necessary to propose a new approach based on nonlinear error growth dynamics for quantifying the predictability of chaotic systems. Ding and Li \cite{DLi} and Li et. al. \cite{LDC} have presented a nonlinear error growth dynamics which applies fully nonlinear growth equations of nonlinear dynamical systems instead of linear approximation to error growth equations to discuss the evolution of initial perturbations and employed it to study predictability. 

\def\a{\delta}

For an $n$-dimensional nonlinear system, the dynamics of small initial perturbation   
${\a}_0={\a}(t_0) \in \mathbb R^n$  about about an initial point  $x_0=x(t_0)$
 in the $n$-dimensional phase space are governed by the nonlinear propagator  
 $\eta(x_0, \a_0, \tau)$, which propagates the initial error forward to the error at the time $t=t_0 + \tau$:  
 $$\a (t_0 + \tau) = \eta(x_0, \a_0, \tau) \a_0.$$
Then the nonlinear local Lyapunov exponent (NLLE) is defined by 
$$\lambda_1(x_0, \a_0, \tau) = \frac{1}{\tau}
\ln\frac{\| \a(t_0 + \tau)\|}{\| \a_0\|}.$$
This indicates that  $\lambda_1(x_0, \a_0, \tau) $ depends generally on the initial state  $x_0$ in the phase space, the initial error  $\a_0$, 
and evolution time  $\tau$. The NLLE is quite different from the global Lyapunov exponent (GLE) or the local Lyapunov exponent (LLE) based on linear error dynamics. If we study the average predictability of the whole system, the whole ensemble mean of the NLLE,  $\bar \lambda_1(\a_0, \tau) =< \lambda_1(x_0, \a_0, \tau) >$ should be introduced, where 
 the symbol $< \cdot > $  denotes the ensemble average.  Then the average predictability limit of a chaotic system could be quantitatively determined using the evolution of the mean relative growth of the  initial error (RGIE) $\bar E(\a_0, \tau) = \exp(\bar \lambda_1(\a_0, \tau)\tau)$.
 According to the chaotic dynamical system theory and probability theory, the saturation theorem of RGIE \cite{DLi}  may be obtained as follows.
 
 \begin{theorem} \cite{DLi} 
 For a chaotic dynamic system, the mean relative growth of initial error (RGIE) will necessarily reach a saturation value in a finite time interval.
 \end{theorem}
 
Once the RGIE reaches the saturation, at the moment almost all predictability of chaotic dynamic systems is lost. Therefore, the predictability limit can be defined as the time at which the RGIE reaches its saturation level.

If the first NLLE   $\lambda_1(x_0, \a_0, \tau)$ along the most rapidly growing direction has been obtained, for an $n$-dimensional nonlinear dynamic system the first $m$ NLLE spectra along other orthogonal directions can be successively determined by the growth rate of the volume  $V_m$ of an $m$-dimensional subspace spanned by the $m$ initial error vectors  $\a_m(t_0)=(\a_1(t_0),  \cdots, \a_m(t_0)$:
 $$\sum^m_{i=1} \lambda_i = \frac{1}{\tau}
\ln\frac{ V_m(\a_m(t_0 + \tau))}{V_m( \a_m(t_0))},$$
for $m=2,3, \cdots, n$, 
where 
$\lambda_i= \lambda_i(x_0, \a_0, \tau) $  is the 
  $i$-th NLLE of the dynamical system. 
  Correspondingly the whole ensemble mean of the $i$-th NLLE is defined as  
  $\bar \lambda_i(\a_0, \tau) =< \lambda_i(x_0, \a_0, \tau) >$.
  
  For a chaotic system, each error vector tends to fall along the local direction of the most rapid growth. Due to the finite precision of computer, the collapse toward a common direction causes the orientation of all error vectors to become indistinguishable. This problem can be overcome by the repeated use of the Gram-Schmidt reorthogonalization (GSR) procedure on the vector frame. Giving a set of error vectors  
  $\{\a_1, \cdots, \a_n\}$, the GSR provides the following orthogonal set  
  $\{\a_1', \cdots, \a_n'\}$:
  \begin{align*}
 & \a_1'=\a_1, \\
 & \a_2'=\a_2 - \frac{(\a_2, \a_2')}{(\a_1', \a_1')}\a_1' , \\
 &\vdots \\
 & \a_n'=\a_n - \frac{(\a_n, \a_{n-1}')}{(\a_{n-1}', \a_{n-1}')}\a_{n-1}' - \cdots -
  \frac{(\a_n, \a_{1}')}{(\a_{1}', \a_{1}')}\a_{1}'.
  \end{align*}
  The growth rate of the $m$-dimensional volume can be calculated by the use of the first $m$ orthogonal error vectors, and then the first $m$ NLLE spectra can be obtained correspondingly. 

On the other hand, we introduce the local ensemble mean of the NLLE in order to measure predictability of specified state with certain initial uncertainties in the phase space and to investigate distribution of predictability limit in the phase space \cite{DLi}. Assuming that all initial perturbations with the amplitude   and random directions are on a spherical surface centered at a initial point  $x_0$, the local ensemble mean of the NLLE relative to   $x_0$  is defined as  
$\bar \lambda_L(x_0, \tau) =< \lambda(x_0, \varepsilon, \tau) >_N$ for $N \to \infty,$
 and then the local average predictability limit of a chaotic system at the point   $x_0$ could be quantitatively determined by examining the evolution of the local mean relative growth of initial error (LRGIE)  
 $\bar E(x_0, \tau) = \exp(\bar \lambda_L(x_0, \tau)\tau)$.
The local ensemble mean of the NLLE different from the whole ensemble mean of the NLLE could show local error growth dynamics of subspace on an attractor in the phase space. Moreover, in practice the local average predictability limit itself might be regarded as a predictand to provide an estimation of accuracy of prediction results.

The nonlinear error growth theory mentioned above provides a new idea for predictability study. However, a great deal of work, including the theory itself, is needed. For a real system such as the atmosphere and ocean, the further studies related to the following questions are needed: 

\begin{enumerate}

\item Quantitative estimates of  the temporal-spatial characteristics of the predictability limit of different variables of the atmosphere and ocean by use of observation data 
(Chen et al. \cite{cld} made a preliminary attempt to this aspect).

\item Relationships among the predictability limits of motion on various time and space scales.

\item Disclosure of the mechanisms influencing predictability from the view of nonlinear error growth dynamics.
\item Predictability limit varying with changes of initial perturbations.

\item Decadal change of the predictability limit.
\item Predictability of extreme events.

\item Prediction of predictability.
\end{enumerate}

\section{Climate Variability and  Successive Bifurcation}
Understanding climate variability and related physical mechanisms and its applications to climate prediction and projection are the primary goals in the study of climate dynamics. One of problems of climate variability research is  to understand and predict the periodic,
quasi-periodic, aperiodic, and fully turbulent characteristics
of large-scale atmospheric and oceanic flows.
Bifurcation theory enables one to determine
how qualitatively different flow regimes appear and
disappear as control parameters vary; it provides us,
therefore, with an important method to explore
the theoretical limits of predicting these flow regimes.

For this purpose,  the ideas of dynamical systems theory and nonlinear functional analysis have been applied so far to climate dynamics mainly by careful numerical
studies. These were pioneered by Lorenz \cite{La, Lb}, Stommel \cite{S},
and Veronis \cite{Va, Vb}
among others, who explored the bifurcation
structure of low-order
models of atmospheric and oceanic flows.

Recently,
pseudo-arclength continuation methods have been applied
to atmospheric (Legras and Ghil \cite{LG}) and oceanic
(Speich et al. \cite{SDG} and Dijkstra \cite{D})
models with increasing horizontal
resolution. These numerical bifurcation studies have produced
so far fairly reliable results for two classes of geophysical flows:
(i) atmospheric flows in a periodic mid-latitude channel, in the presence of
bottom topography and a forcing jet; and (ii) oceanic
flows in a rectangular mid-latitude basin, subject to
wind stress on its upper surface; see among others Charney and DeVore \cite{CD},
Pedlosky \cite{Pa}, Legras and Ghil \cite{LG} and Jin and Ghil \cite{JG}  for saddle-node and Hopf bifurcations in the the atmospheric channel, and
\cite{BM, CI, IS, JJG, MB, SDG, CGIL,SGITWa, SGITWb, M, NL}  for saddle-node, pitchfork or Hopf or global bifurcation 
in the oceanic basin. 

Apparently, further numerical bifurcation studies are inevitably necessary.
Typical problems include 1)  continuation algorithms (pseudo-arclength methods and  stability analysis) applied to large-dimensional dynamical systems (discretized PDEs), 
2)  Galerkin approach using finite-element discretization together with 
`homotopic' meshes that can deform continuously from a domain to another. 

Another important further direction is  to rigorously conduct 
bifurcation and stability analysis for the original partial differential equations models associated with typical phenomena.
Some progresses have been made in this direction; see among others 
\cite{hmw1,hmw2,hmw3, cgsw}. It is clear that much more effort is needed; see also 
Section 6 below.

Furthermore, very little is known for theoretical and numerical investigations on the  bifurcations of coupled systems, which are of practical significance for the coupled dynamics. It is also practically important to study the variability and dynamics of  systems under varying external forcing, e.g., the features, processes and dynamics of weather and climate varies with the global warming.

Hereafter in next few subsections, we present some issues on a few specific physical phenomena. 

\subsection{Wind-driven and thermohaline circulations}
 For the ocean, 
basin-scale motion is dominated by wind-driven (horizontal) 
and thermohaline (vertical) circulations. 
Their variability, independently and interactively, may play a significant role 
in climate changes, past and future. The wind-driven circulation 
plays a role mostly in the oceans' subannual-to-interannual variability, 
while the thermohaline circulation is most important in decadal-to-millenial 
variability.

 The thermohaline circulation (THC) is highly nonlinear due to the combined effects of the temperature and  the salinity on density (Meinckel et. at. \cite{MQBB}; Rahmstorf \cite{rahm}), which cause the existence of multiple equilibria and thresholds in the THC. The abrupt climate is related to the shift between multiple equilibria flow regimes in the THC. The sensitivity of the THC to anthropogenic climate forcing is still an open question (Rahmstorf  \cite{rahm}; Meincke  et. al. \cite{MQBB}; Thorpe et al., \cite{TGJWM}). 
THis is closely related to the question on  whether an abrupt breakdown of the THC can result from global warming.
In particular, there are two different but connected the stability and transitions associated with the problem. The first is  stability and transitions of  the solutions of the partial differential equation models in the phase space,  and the second is the structure of the solutions and its transitions in the physical spaces. Issues related to these transitions appear to be very important. 
One such example is the western boundary current separation.
The physics of the separation of western boundary currents   
is a long standing problem in physical oceanography.   
The Gulf Stream in the North Atlantic and Kuroshio
in the North Pacific   have a fairly  similar behavior 
with separation from the coast 
occurring at or close to a fixed latitude. 
The Agulhas Current in the Indian Ocean, however, 
behaves  differently by showing a retroflection accompanied  
by ring formation. The 
current rushes southward along the east coast of the African, 
overshoots the southern 
latitude of this continent and then suddenly it turns eastward 
and flows backward into the 
Indian Ocean.   The North Brazil  Current in the equatorial  
Atlantic shows a  similar, but weaker, retroflection.

Mathematically speaking, the  boundary-layer separation problem is crucial for understanding the transition to turbulence and stability properties of fluid flows. 
This problem  is also closely linked to
structural and dynamical bifurcation of the flow through 
a topological change of its
spatial and  phase-space structure. This program of research has been initiated by 
Tian Ma and one of the authors in this article, in collaboration in part with 
Michael Ghil; see \cite{amsbook, mw01a,mw02b,mw03c,mw03e,mw04,gmw1, gmw2,glww}. A great deal of further studies in this direction are needed.

\subsection{Intraseasonal Oscillations (ISOs)} 
Another important source of variability is related to ISOs 
such as the Madden-Julian Oscillation (MJO). MJO  is a large-scale oscillation (wave) 
in the equatorial region (Madden and Julian \cite{MJ71,MJ72}) 
and is  the dominant component of the intraseasonal 
(30-90 days) variability in the tropical atmosphere (Zhang \cite{zhang05}). 
Although some theories and hypotheses have been proposed to understand 
the MJO, a completely satisfactory dynamical theory for the MJO has not 
yet been established (Holton \cite{holden}).

The basic governing equations used to theoretically analyze and simulate the MJO could be found in Wang \cite{wangbin}. From the mathematical point of view, 
the well-posedness, asymptotic behavior of the equations, and  bifurcation and
stability analysis are the first theoretical questions to be answered.

The observation diagnosis indicates that there is a rich 
multiscale structure of the MJO, and scale interactions 
might play an important role in the MJO (Slingo et al. \cite{SINWY}). 
However, whether the scale interactions are essential for the 
scale selection of the MJO is an important open question 
(Wang \cite{wangbin, zhang05}). It is therefore  necessary to 
develop a multiscale analysis theory of the multiscale model to study the upscale energy transfer and 
to recognize the formation of large-scale structure of the system through 
multiscale interactions. Two basic aspects might be involved into 
this area. One is the significance of short-term cycle in the 
life cycle of the inherent large-scale structure of a dynamical system, 
e.g., the diurnal cycle to the MJO. The other is how the mesoscale and 
synoptic-scale systems go through certain organized action or 
stochastic dynamics to form a massive behavior and influence 
movement of this large-scale structure.

Recently, extratropical ISO or mid-high latitude low-frequency variability (LFV) has been revealed, e.g., the 70-day oscillation found over the North Atlantic (Felik, Gihl and Simonnet, \cite{FGS}; Keppenne,  Marcus,  Kimoto and Ghil \cite{KMKG}; 
Lau, Sheu and Kang \cite{LSK}) that is related to LFV of North Atlantic Oscillation (NAO). The dynamics of extratropical ISO or LFV is clearly an attractive field in the future. 

\subsection{ENSO}
The El Ni\~no-Southern Oscillation (ENSO) is the known strongest interannual climate variability associated with strong atmosphere-ocean coupling, which has significant impacts on global climate. ENSO is in fact a phenomenon that warm events (El Ni\~no phase) and clod events (La Ni\~na phase) in the equatorial eastern Pacific SST anomaly occur by turns, which associated with persistent weakening or strengthening in the trade winds by turns.

It is convenient, effective, and easily understandable to employ the simplified coupled dynamical models to investigate some essential behaviors of ENSO dynamics. The simplest and leading theoretical models for ENSO are the Òdelayed oscillatorÓ model (Schopf and Suarez \cite{SS},  Battisti and Hirst \cite{BH}) and the Òrecharge oscillatorÓ model (Jin  \cite{jin}). However, the basic shortcoming of those highly simplified models cannot account for the observed irregularity of ENSO, although they could qualitatively explain the average features of an ENSO cycle. Hence further study  is inevitably necessary.
Another simplified coupled ocean-atmosphere model which can be used to predict ENSO event is the Zebiak and Cane (ZC) model (Zebiak and Cane \cite{ZC}). The atmosphere model is the Gill-type (Gill \cite{gill80}; Neelin et al. \cite{neelin}), and the ocean model consists of a shallow-water layer with an embedded mixed layer.
Also, we would like to mention the intriguing behavior of Boolean Delay Equations (BDE) in the ENSO context; see  Ghil,  Zaliapin and Coluzzi \cite{GZC}  and the references therein. It is worth studying the dynamical bifurcation and stability of solutions of this kind of simplified models to understand the phase transitions of ENSO and its low-frequency variability; see also the dynamical bifurcation theory in the next section.%This is a coupled bifurcation and coupled stability problem of solutions.

The ENSO coupling processes and dynamics under the global warming by using successive bifurcation theory and the predictability of ENSO by using of nonlinear error growth theory are also of considerable practical importance.
However, an interesting current debate is whether ENSO is best modeled as a stochastic or chaotic system - linear and noise-forced, or nonlinear oscillatory and unstable? It is obvious that a careful fundamental level examination of the problem is crucial.

\section{New Dynamical Systems Theories and Geophysical Applications} 

\subsection{Introduction and motivation}
As mentioned earlier, most studies on bifurcation issues in geophysical fluid dynamics so far have only considered systems of ordinary
differential equations (ODEs) that are obtained by projecting
the PDEs onto a finite-dimensional solution space, either
by finite differencing or by truncating a Fourier expansion
(see Ghil and Childress \cite{GC} and further references there).

A challenge mathematical problem is to conduct rigorous bifurcation and 
stability analysis for the original partial differential equations (PDEs)
that govern geophysical flows. 
Progresses in this area should allow us to overcome some of 
the inherent limitations of the numerical
bifurcation results that dominate the climate dynamics literature up
to this point,  and
to capture the essential dynamics of
the governing PDE systems.

Recently, Ma and Wang initiated a study on a new dynamic bifurcation and 
stability theory for dynamical systems. 
This bifurcation theory is centered at a new notion of bifurcation, called 
attractor bifurcation for dynamical systems, both finite dimensional and 
infinite dimensional. The main ingredients of the theory include 
a) the attractor bifurcation theory, b) steady state bifurcation for a 
class of nonlinear problems with even order non-degenerate 
nonlinearities, regardless of the multiplicity of the 
eigenvalues, and c) new strategies for 
the Lyapunov-Schmidt reduction and the center manifold 
reduction procedures. The general philosophy is that we first derive general existence 
of bifurcation to attractors, and then we classify the bifurcated 
attractors to derive detailed 
dynamics including for instance stability
of the bifurcated solutions.

The bifurcation theory has been applied to 
various problems from science and engineering, including, in particular, 
the Kuramoto-Sivashinshy equation, the Cahn-Hillard equation, 
the Ginzburg-Landau equation, Reaction-Diffusion equations 
in Biology and Chemistry, and the B\'enard convection problem and the Taylor problem in classical fluid mechanics; see a recent book by Ma and Wang \cite{b-book}. 
For applications to geophysical fluid dynamics problems, we have carried out the detailed bifurcation and stability analysis for 1) the stratified Boussinesq equations \cite{hmw3}, 
2) the doubly-diffusive modes (both 2D and 3D) \cite{hmw1, hmw2}. 
%In Section~\ref{s6.3} below, we shall give a brief description of the results for the 
%stratified Boussinesq equations based on \cite{hmw3}.

\begin{figure}
        \centering \includegraphics[height=0.4\hsize]{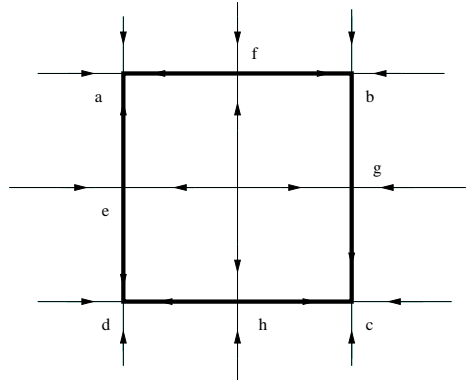}
        \caption{A bifurcated attractor containing 4 nodes (the points a, b, c, and d), 4 saddles (the points e, f, g, h), and orbits connecting these $8$ points.}
\label{fg4.1}
\end{figure}

We proceed with a simple example to illustrate the basic motivation and ideas behind the attractor bifurcation theory.
For $x =(x_1, x_2) \in \R^2$, the system $\dot x = \lambda x - (x_1^3, 
x_2^3) + o(|x|^3)$
bifurcates from $(x, \lambda)=(0,0)$ to an attractor  $\Sigma_\lambda=S^1$.
This bifurcated attractor is as shown in Figure \ref{fg4.1}, and contains 
exactly 4 nodes (the points a, b, c, and d), 4 saddles (the points e, f, 
g, h), and orbits connecting these $8$ points. 
From the physical transition point of view, as $\lambda$ crosses $0$, the new state after the system undergoes a transition is represented by the {\bf whole bifurcated attractor} $\Sigma_\lambda$, rather than any of the steady states or any of the connecting orbits.
The connecting orbits represents transient states. 
Note that the global attractor  is the 2D region enclosed by $\Sigma_\lambda$.
We point out here that the bifurcated attractor is different from the study on global attractors of a dissipative dynamical system-both finite and infinite dimensional. Global attractor studies the global long time dynamics (see among others \cite{FT79, BV83, CF85, CFT85}), while the bifurcated attractor provides a natural object for studying dynamical transitions \cite{b-book, chinese-book, ptd}.

One important characteristic of the new attractor bifurcation theory 
is related to the asymptotic stability of the bifurcated solutions. 
This characteristic can be viewed as follows. First, 
as an attractor itself, the  bifurcated attractor has a basin of attractor, 
and consequently is a useful object to describe local transitions.
Second, with detailed classification of the solutions in 
the bifurcated attractor,  we are able to access not only 
the asymptotic stability of the bifurcated attractor, 
but also the stability of 
different solutions in the bifurcated attractor, providing a more complete
understanding of the transitions of the physical system as the system 
parameter varies. 
Third, as Kirchg\"assner \cite{kirch} indicated, 
an ideal stability theorem would include all physically meaningful 
perturbations and today we are still far from this goal.
In addition, fluid flows are normally time dependent. Therefore  
bifurcation analysis for steady state problems provides in general 
only partial answers to the problem, and is not enough for solving 
the stability problem. 
Hence from the physical point of view, 
attractor bifurcation provides a nature tool for studying transitions for deterministic systems.

\subsection{A brief account of the attractor bifurcation theory}
We now briefly present this new attractor bifurcation theory and  refer the interested readers to \cite{b-book, chinese-book} for details of the theory and its various applications.

We start with a basic state $\bar \varphi$ 
of the system, a steady state solution of (\ref{3.1}).
Then consider  $\varphi=u+ \bar \varphi$.
Then (\ref{3.1}) becomes 
\begin{align}
\label{4.1}
& \frac{du}{dt} = L_\lambda u + G(u,\lambda),\\
\label{4.2}
&u(0) = u_0,
\end{align}
where $H$ and $H_1$ are two Hilbert spaces such that 
$H_1\to H$ be a dense and compact inclusion, 

The mapping $u: [0, \infty) \to H$  is the unknown function, 
$\lambda \in \R$  is the  system  parameter, and 
$L_\lambda :H_1\to H$ is a family of linear completely continuous fields
depending continuously on $\lambda \in \R$, such that
\begin{equation}
\label{4.3}
\left.  \begin{aligned}
&L_\lambda =-A+B_\lambda && \quad \text{a sectorial operator,}\\
&A:H_1 \to H && \quad \text{a linear homeomorphism,} \\
&B_\lambda :H_1 \to H && \quad \text{a linear compact operator.}
\end{aligned} \right.
\end{equation}
It is known that $L_\lambda$ generates an analytic semigroup
$\{e^{-tL_\lambda}\}_{t\ge 0}$ and we can define fractional power
operators $L^\alpha_\lambda$ for $\alpha \in \R$ with domain $H_\alpha
=D(L_\lambda^\alpha)$ such that $H_{\alpha_1} \subset H_{\alpha_2}$
is compact if $\alpha_1>\alpha_2$, $H=H_0$ and $H_1 = H_{\alpha =1}$.

Furthermore, we assume that for some $\theta <1$ the nonlinear
operator $G(\cdot,\lambda):H_\theta \to H_0$ is a $C^r$ bounded
operator $(r\ge 1)$, and 
\begin{equation}
\label{4.4}
G(u,\lambda) = o(\|u\|_\theta), \qquad \forall\,\, \lambda \in \R.
\end{equation}

\begin{definition}[Ma \& Wang \cite{mw-db1,b-book}]
\label{df3.6}
\begin{enumerate}

\item We say that the equation (\ref{4.1}) bifurcates from
$(u,\lambda) = (0,\lambda_0)$ to an invariant set $\Sigma_\lambda$, if
there exists a sequence of invariant sets $\{\Sigma_{\lambda_n}\}$ of
(\ref{4.1}), such that  $0 \notin \Sigma_{\lambda_n}$,  and 
$$\lim_{n\to \infty} \lambda_n = \lambda_0, \qquad \lim_{n\to \infty} \max_{x\in \Sigma_{\lambda_n}} \|x\| =0.
$$

\item If the invariant sets $\Sigma_\lambda$ are attractors of
(\ref{4.1}), then the bifurcation is called an attractor
bifurcation. 

\end{enumerate}
\end{definition}
\begin{figure}
        \centering \includegraphics[height=0.5\hsize]{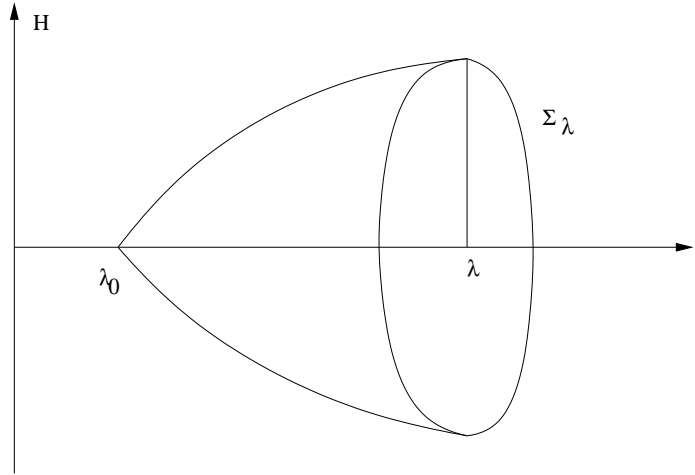}
        \caption{Definition of attractor bifurcation.}
\label{fg1}
\end{figure}

Let $\{\beta_k(\lambda) \in \C \mid k=1,2,\cdots\}$ be the
eigenvalues of $L_\lambda = -A+B_\lambda$ (counting the multiplicities).
Suppose that 
\begin{align}
\label{4.5}
& 
Re\beta_i (\lambda) 
\left\{ 
\begin{aligned}
&<0 \quad && \text{ if } \lambda<\lambda_0 \\
&=0 \quad && \text{ if } \lambda=\lambda_0 \\
&>0 \quad && \text{ if } \lambda>\lambda_0
\end{aligned} \right.
&& \text{for $1\le i\le m$,} \\
\label{4.6}
& Re\beta_j(\lambda_0) <0 \qquad  &&\text{for $j\ge m+1$.}
\end{align}
Let$E_0$ be the eigenspace of $L_\lambda$ at $\lambda_0$ 
\[
E_0 = \bigcup_{1\le i\le m} \left\{u\in H \mid (L_{\lambda_0} -
\beta_i (\lambda_0))^k u=0, \, k\in \N\right\}.
\]
By (\ref{4.5}) we know that $\dim E_0 =m$.

In physical terms, the above properties of eigenvalues are  called principle of exchange of stabilities (PES).  They can be verified in many physical systems. It is obvious that they are necessary for linear instability and the following attractor bifurcation theorem demonstrates  that they lead to bifurcation to an attractor.

\begin{theorem}[Ma  \& Wang \cite{b-book,mw-db1}]
\label{th3.7}
Assume that (\ref{4.5}) and (\ref{4.6}) hold true, and let $u=0$ be
a locally asymptotically stable equilibrium point of (\ref{4.1}) at
$\lambda=\lambda_0$. Then we have the following assertions.
\begin{enumerate}
\item Equation (\ref{4.1}) bifurcates from $(u,\lambda) =
(0,\lambda_0)$ to an attractor $\Sigma_\lambda$ for $\lambda>\lambda_0$
with $m-1\le \dim \Sigma_\lambda \le m$, which is an
$(m-1)$-dimensional homological sphere.

\item For any $u_\lambda \in \Sigma_\lambda$, $u_\lambda$ can be
expressed as 
\[
u_\lambda = v_\lambda + o(\|v_\lambda\|), 
\]
where $v_\lambda \in E_0$.

\item There exists a neighborhood $U \subset H$ of $u=0$ such that
$\Sigma_\lambda$ attracts $U\setminus \Gamma$, where $\Gamma$ is the stable
manifold of $u=0$ with codimension~$m$. In particular, if $u=0$ is
global asymptotically stable for (\ref{4.1}) at $\lambda=
\lambda_0$ and (\ref{4.1}) has a global attractor for any $\lambda$
near $\lambda_0$, then $\Sigma_\lambda$ attracts $H\setminus \Gamma$.
\end{enumerate}
\end{theorem}

With this general theorem, along with other important ingredients addressed in Ma and Wang \cite{b-book, chinese-book},  in our disposal, the main strategy to  conduct the bifurcation analysis  consists of 1) the existence of attractor bifurcation, and 2) classification of solutions in the bifurcated attractor. We refer the interested readers to the books \cite{b-book, chinese-book} for details; see also the discussion below on the Rayleigh-B\'enard convection.

\subsection{Dynamic bifurcation in classical fluid dynamics}%for the Rayleigh-B\'enard convection problem}
%\subsection{Stratified Boussinesq equations}\label{s6.3}
The new dynamic bifurcation theory has  been naturally applied to problems in fluid dynamics, including the 
Rayleigh-B\'enard convection, the Taylor problem, and the parallel 
shear flows, by Ma and Wang and their collaborators. 
The study of these basic problems on the one hand plays an important role in 
understanding the turbulent behavior of fluid flows, 
and on the other hand often leads to new insights and methods toward 
solutions of other problems in sciences and engineering. 

To illustrate the main ideas of the applications of the theory,  we consider now  the Rayleigh-B\'enard convection problem. 
Linear theory of the Rayleigh-B\'enard problem were essentially derived by physicists; see, among others, Chandrasekhar \cite{chandrasekhar} and  Drazin and Reid \cite{dr}.
Bifurcating solutions of the nonlinear problem were first constructed formally 
by Malkus and Veronis \cite{veronis58}. The first rigorous proofs of the existence of 
bifurcating solutions were given by Yudovich \cite{yudovich67a, yudovich67b} and 
Rabinowitz \cite{rabinowitz}. Yudovich  
proved the existence of bifurcating solutions by a topological degree argument. Earlier, however, Velte \cite{velte} had proved the existence of branching 
solutions of the Taylor problem by a topological degree argument as well. 
The have been many studies since the above mentioned early works.  However, as we shall see, a rigorous and complete understanding of the problem is not available until the recent work by Ma and Wang using the attractor bifurcation theory. 
\def\pr{{P_r}}
\def\sPr{\sqrt{\pr}}
\def\R{\mathbb R}
\def\mA{\mathcal A}
\def\sR{\sqrt{R}}
\def\p{\partial}
\def\tilde{\widetilde}

To illustrate the ideas, we
 start by recalling the basic set-up of the problem. 
Let   the  Rayleigh number  be  
$$R=\frac{g\alpha (\bar{T}_0-\bar{T}_1)h^3}{(\kappa\nu)}.$$
The B\'enard convection is modeled by the Boussinesq equations. In their nondimensional form, these equations are written as follows:
\begin{equation}
\label{eq2.4}
\left.
\begin{aligned}
& \frac{1}{\pr} 
   \left[\frac{\partial u}{\partial t} + (u \cdot \nabla ) u + \nabla p\right]
 -  \Delta u -  \sR T k =0, \\
& \frac{\partial T}{\partial t} + (u\cdot \nabla )T - \sR u_3 - \Delta T =0, \\
& \divv\,\, u = 0.
\end{aligned}
\right.
\end{equation}
Here the unknown functions are $(u, T, p)$, which are the deviations from the basic field. The non-dimensional domain is $\Omega =
D\times (0,1) \subset \R^3$, where $D\subset \R^2$ is an open set. The 
coordinate system is given by  $x=(x_1,x_2,x_3) \in \R^3$.
They are supplemented with the following  initial value
conditions 
\begin{equation}
\label{eq2.7}
(u, T) = (u_0, T_0) \qquad \text{ at } t=0.
\end{equation}

Boundary conditions are needed at the top and bottom  
and at the lateral boundary $\partial D \times (0, 1)$. 
At the top and bottom boundary ($x_3=0, 1$), either the so-called rigid or 
free boundary conditions are given
\begin{equation}
\label{eq2.8}
\left.
\begin{aligned}
& T =0, \quad u =0 &&  \text{(rigid boundary)}, \\
& T=0, \quad u_3 =0,  \quad \frac{\p (u_1, u_2)}{\p x_3} =0 
&&  \text{(free boundary)}.
\end{aligned}\right.
\end{equation}

Different combinations of top and bottom boundary conditions are normally 
used in different physical setting such as {\it rigid-rigid}, 
{\it rigid-free}, {\it free-rigid}, and {\it free-free}.

On the lateral boundary $\partial  D\times [0,1]$, one of the following 
boundary conditions are usually used:

\begin{enumerate}

\item Periodic condition:
\begin{equation}
\label{eq2.10}
(u, T)(x_1+ k_1 L_1, x_2+ k_2 L_2, x_3)=(u, T)(x_1, x_2, x_3), 
\end{equation}
for any $k_1, k_2 \in \mathbb Z$.

\item Dirichlet boundary condition:
\begin{equation}
\label{eq2.11}
u=0, \quad T=0 \quad \text{(or $\frac{\p T}{\p n}=0$)};
\end{equation}

\item Free boundary condition:
\begin{equation}
\label{eq2.12}
T=0, \quad u_n =0, \quad \frac{\partial u_\tau}{\partial  n} = 0,
\end{equation}
where $n$ and $\tau$ are the unit normal and tangent vectors on
$\partial D\times [0,1]$ respectively, and $u_n = u\cdot n$, $u_\tau
= u\cdot \tau$.
\end{enumerate}

By using the attractor bifurcation theory, the following results have been obtained by Ma and Wang in \cite{mw-benard, mw07, b-book, chinese-book}.

\begin{itemize}

\item[(1)] When the
Rayleigh number $R$ crosses the first critical Rayleigh number $R_c$, the Rayleigh-B\'enard  problem bifurcates from the basic state to an  attractor $A_R$, homologic to $S^{m-1}$,
where $m$ is the multiplicity of $R_c$  as an eigenvalue of the linearized problem near the basic solution, for all physically  sound boundary conditions, regardless of  the geometry of the domain and the  multiplicity of the eigenvalue $R_c$ for the linear problem

\item[(2)] Consider the 3D B\'enard convection in 
$\Omega=(0, L_1)\times (0, L_2) \times (0, 1)$  with free top-bottom and periodic horizontal boundary conditions, and with  
\begin{equation}
\frac{k^2_1}{L^2_1}+\frac{k^2_2}{L^2_2}=\frac{1}{8} \qquad \text{ for some }
k_1,k_2\in\mathbb{Z}.\label{9.58}
\end{equation}
Then 
\begin{equation}
A_R=\left\{
\begin{aligned}
& S^5  && \qquad  \text{if}\ \ \ \ L_2=\sqrt{k^2-1}L_1,\ \
\ \ k=2,3,\cdots ,\\
& S^3 &&\qquad  \text{otherwise}.
\end{aligned}\right.
\end{equation}

\item[(3)] For the 3D B\'enard convection   in $\Omega=(0, L)^2 \times (0, 1)$   with free boundary conditions and  with 
\begin{equation}
0<L^2<\frac{2-2^{{1}/{3}}}{2^{{1}/{3}}-1}.\label{9.68}
\end{equation}
The  bifurcated  attractor $A_R$ consists of   
   exactly eight singular
points  and   eight heteroclinic orbits connecting the singular points, as shown in Figure \ref{fg4.1}, 
with 4 of them  being  minimal attractors, and the other 4
  saddle points.

\end{itemize}

The proof of these results is carried out in the following steps:

\begin{itemize}

\item[1)] It is classical to put the Boussinesq equations (\ref{eq2.4}) in the abstract form  (\ref{4.1}). Then it is easy to see that the linearized operator is a self-adjoint operator., and both conditions  (\ref{4.3}) and (\ref{4.4}) are satisfied.

\item[2)] As the linearized operator is symmetric, it is not hard to verify the PES  (\ref{4.5}) and (\ref{4.6}) for the Boussinesq equations.

\item[3)] To derive a general attractor theorem for the Benard convection regardless of the domain and the boundary conditions, we need to verify the asymptotic stability of the basic solution $(u, T)=0$ at the critical Reynolds number. This can be achieved by a general stability theorem, derived by Ma and Wang in \cite{mw-benard}. 

\item[4)] The detailed structure of the solutions in the bifurcated attractor can be derived using a new approximation formula for the center manifold function; see Ma and Wang \cite{mw07, b-book, chinese-book}.

\end{itemize}

We remark here that the high dimensional sphere $S^{m-1}$ contains not only the steady states generated by symmetry groups inherited in the problem, but also many transient states, which were completely missed by any classical theories. These transient states are highly relevant in geophysical fluid dynamics and climate dynamics. We believe the climate  low frequency climate variabilities discussed in the previous section are related to certain transient states, and further exploration in this direction are certainly important and necessary.

\subsection{Dynamic bifurcation and stability in  geophysical fluid dynamics} 
As mentioned earlier, the theory has been applied to models in geophysical fluid dynamics models including the doubly-diffusive models, and rotating Boussinesq equations. We now briefly address these applications in turn.

{\sc Rotating Boussinesq Equations:} Rotating Boussinesq equations are basic model in atmosphere and ocean dynamical models. In \cite{hmw3},  for the case where the Prandtl
number is greater than one, a complete stability and bifurcation
analysis near the first critical Rayleigh number is carried out.
Second, for the case where the Prandtl number is smaller than one,
the onset of the Hopf bifurcation near the first critical Rayleigh
number is established, leading to the existence of nontrivial
periodic solutions. The analysis is based on a newly developed bifurcation and stability theory for nonlinear dynamical systems  as mentioned above.

{\sc Double-Diffusive Ocean Model:} 
Double-diffusion was first originally discovered 
in the 1857 by Jevons \cite{jevons}, forgotten, and then rediscovered
as an ``oceanographic curiosity'' a century later; see  among others 
Stommel, Arons and 
Blanchard \cite{stommel}, Veronis \cite{gv}, 
and Baines and Gill \cite{pg}. In addition to its 
effects on oceanic circulation, double-diffusion 
convection has wide applications to  such diverse fields 
as growing crystals, the dynamics of magma chambers 
and convection in the sun.

The best known doubly-diffusive instabilities are ``salt-fingers'' 
as discussed in the pioneering work by Stern \cite{stern}. 
These arise when hot salty water lies over cold fresh 
water of a higher density and consist of long fingers of rising and 
sinking water. A blob of hot salty water which finds itself surrounded 
by cold fresh water rapidly loses its heat while retaining its salt due 
to the very different rates of diffusion of heat and salt. The blob 
becomes cold and salty and hence denser than the surrounding fluid. 
This tends to make the blob sink further, drawing down more hot 
salty water from above giving rise to sinking fingers of fluid. 

In \cite{hmw1, hmw2}, we present a bifurcation and stability analysis on 
the doubly-diffusive convection. 
The main objective  is to study  1) the mechanism of 
the saddle-node bifurcation  and hysteresis for the problem, 
2) the formation, stability and transitions of the typical convection 
structures,  and 3) the stability of  solutions.
It is proved in particular  that there
are two different types of transitions: continuous and jump, 
which are determined explicitly using some physical relevant 
nondimensional parameters. It is also proved that the jump 
transition always leads to the existence of a saddle-node 
bifurcation and hysteresis phenomena.

However, there are many issues still open, including in particular to use the dynamical systems tools developed to study some of the issues raised in the previous sections.

\subsection{Stability and transitions of geophysical flows in the physical space}
Another important area of studies in geophysical fluid dynamics is to study the structure and its stability and transitions of  flows in the physical spaces. 

A method to study these important problems in geophysical fluid dynamics is a recently developed  geometric theory for incompressible flows by Ma and Wang  \cite{amsbook}.
This theory   consists of research in directions: 
1) the study of the structure and its transitions/evolutions 
of divergence-free vector fields, and 2) the study of 
the structure and its transitions 
of velocity fields for 2-D incompressible fluid flows governed 
by the Navier-Stokes equations or the Euler equations.
The study in the first direction  is 
more kinematic in nature, and the results and methods 
developed can naturally be applied to other problems of mathematical physics 
involving divergence-free vector fields.
In fluid dynamics context, the study in the second direction 
involves specific 
connections between the solutions of the Navier-Stokes or the Euler 
equations and flow structure in the physical space. In other words, 
this area of research links the kinematics to the dynamics of 
fluid flows. This is unquestionably an important and difficult problem.

Progresses have been made in several directions. First, 
a new rigorous characterization of 
boundary layer separations for 2-D viscous incompressible flows is developed 
recently by Ma and Wang, in collaboration in part 
with Michael Ghil \cite{amsbook}.
The nature of flow's boundary layer separation from the 
boundary plays a fundamental role in many physical 
problems, and often determines the nature of the flow in the 
interior as well. The main objective of this section is to 
present a  rigorous characterization of 
the boundary layer separations of 2D incompressible
fluid flows. This is a long standing problem in fluid 
mechanics going back to the pioneering work of Prandtl \cite{prandtl} 
in 1904. 
No known theorem, which can be applied to determine the separation, 
is available until the recent work by Ghil, Ma and Wang \cite{gmw1, gmw2}, and 
Ma and Wang \cite{mw01a, mw03c}, 
which provides a first rigorous characterization. 
Interior separations are studied rigorously by Ma and Wang \cite{mw04}. The results for both the interior and boundary layer separations are used in Ma and Wang \cite{mw-tcp}  for the transitions of the Couette-Poiseuille  and the Taylor-Couette-Poiseuille flows.

Another example in this area is the justification 
of the roll structure (e.g. rolls) in the 
physical space in the Rayleigh-B\'enard convection by Ma and Wang \cite{mw-benard, mw07}. We note that  a special  structure with rolls separated by a cross channel flow derived 
in \cite{mw07} has not been rigorously examined in the B\'enard convection setting although it has been  observed in other physical contexts such as the Branstator-Kushnir waves in the atmospheric dynamics 
\cite{branstator,kushnir}.

With this theory in our disposal, 
the structure/patterns and their stability and transitions 
in the underlying physical spaces for those problems in fluid dynamics 
and in geophysical fluid dynamics can be classified. 
In particular, this theory has been used to  study the  formation, persistence and transitions of flow structures including 
boundary layer separation including the Gulf separation, the Hadley circulation, and the Walker circulation. Further investigation appears to be utterly important and necessary. 
In addition, it appears that such theoretical and numerical studies will lead to better predications on weather and climate regimes.

%\nocite{whc,LC03,LC01,LC99a,LC99b,LC98a,LC98b,LC97a,LC97b,LC97c,
%LC96a,LC96b,LC95}
\bibliography{li-wang}

\end{document}